\documentclass{article}
\usepackage{spconf,amsmath,epsfig}
\usepackage{xcolor}
\usepackage[
singlelinecheck=false % <-- important
]{caption}

\pdfoutput=1

\title{Time-Space tradeoff in deep learning models for crop classification on satellite multi-spectral image time series}

\name{V. Sainte Fare Garnot$^{1}$, L. Landrieu$^{1}$, S. Giordano$^{1}$, N. Chehata$^{1,2}$}
\address{$^{1}$  Univ. Paris-Est, LASTIG-STRUDEL, IGN-ENSG, F-94160 Saint-Mand\'{e}, France\\
$^{2}$  EA G\&E Bordeaux INP, Universit\'e Bordeaux Montaigne, France\\
                }

\begin{document}
%\ninept
%
\maketitle
\begin{abstract}
In this article, we investigate several structured deep learning models for crop type classification on multi-spectral time series. In particular, our aim is to assess the respective importance of spatial and temporal structures in such data. With this objective, we consider several designs of convolutional, recurrent, and hybrid neural networks, and assess their performance on a large dataset of freely available Sentinel-2 imagery. We find that the best-performing approaches are hybrid configurations for which most of the parameters (up to 90\%) are allocated to modeling the temporal structure of the data. Our results thus constitute a set of guidelines for the design of bespoke deep learning models for crop type classification. 
\end{abstract}
\section{Introduction}
\label{sec:intro}
Agricultural crop type mapping provides a highly valuable tool for agricultural management, environmental monitoring, or policy making. With the development of publicly available medium-high resolution remote sensing data such as the multi-spectral image time series provided by the Sentinel-2 mission, crop classification can now be implemented on a large scale and thus calls for increasingly high-performing automated machine-learning based classification tools.

In order to achieve this, discriminative models such as Random Forest (RF) classifiers have been extensively used for their good performance and the easy interpretability of their classification decisions \cite{inglada2015assessment,singh2018crop,bailly2018crop}. 
Yet, such algorithms fail to take advantage of the spatial and temporal structures of the data at hand. Consequently, several works attempt to explicitly model the temporal structures: modelling multi-annual crop rotation using Conditional Random Fields \cite{bailly2018crop}, or modelling phenological properties with Hidden Markov Models \cite{siachalou2015hidden} for instance.
More recently, advances in deep learning provided very efficient tools for complex data analysis. Specifically,  Convolutional Neural Networks (CNN) \cite{lecun1995convolutional} and Recurrent Neural Networks (RNN) \cite{hochreiter1997long} are proving very efficient at exploiting spatial and temporal structure respectively. Additionally, RNNs and CNNs can be combined in hybrid recurrent convolutional networks (RCNN) in order to exploit both spatial and temporal structures in time series of images. These methods are being successfully applied for crop type mapping in a variety of manners: using CNN for unitemporal and multi-temporal classification \cite{ji20183d}, RNN to model crop phenology \cite{russwurm2017multi}, or RCNN for multi-temporal crop segmentation \cite{russwurm2018multi}.

In this article, we consider the problem of crop classification on multi-spectral time series, assuming that parcel segmentation is already known. We attempt  to answer the following question: Given a fixed budget of parameters, should one prioritize modeling the spatial structure (with CNN), temporal structure (with RNN), or both (with RCNN)? We compare the crop classification performance of several architectures with the same number of parameters on a Sentinel-2 dataset of agricultural parcels.

The key contributions of our work are as follows:
\begin{itemize}
\vspace{-0.3cm}
    \item We provide an empirical analysis of the features extracted by convolutional and recurrent models.
    \vspace{-0.3cm}
    \item We show that recurrent architectures are acting as a memory combining multiple observations, as well as a model for temporal evolution.
    \vspace{-0.3cm}
    \item We provide empirical evidence that the temporal structure of Sentinel-2 data is richer than the spatial structure for crop type classification, and, consequently, that most of a model's complexity should be allocated to it.
\end{itemize}

\section{Methods}
\label{sec:meth}

\subsection{Neural Network Architectures}
In order to assess the influence of the temporal and spatial structure, we implement four neural network  architectures. All of them share the same scheme: first extract an embedding  --- spatial, temporal or both --- from the input image sequence and then perform the classification based on this embedding. The same design of classifier unit is used across all architectures: a Multi Layer Perceptron (MLP) with two hidden layers of dimension 128 and 64.
%\vspace{-.6cm}
\subsubsection{CNN}
We first implement a convolution-based neural network whose goal is to leverage the spatial structure of image time series. The images corresponding to each date are embedded independently through the same three layers composed of the following units:  convolution with $3\times3$ kernel size and no pad\-ding, batch normalization \cite{ioffe2015batch}, ReLu activation \cite{glorot2011deep} and MaxPool with $2\times2$ kernel size. We then compute a global embedding for the whole sequence by concatenating all the image embeddings and taking the maximum for each channel, in the manner of PointNet \cite{qi2017pointnet}. Finally, the global embedding is fed to the MLP for classification. 

\vspace{-.2cm}
\subsubsection{RNN}
Unlike the CNN network, the RNN architecture focuses purely on the temporal dimension of image sequences. 
For each image, we compute a vector of parcel-level handcrafted features: the mean and standard deviation of each spectral band.
These vectors are then fed in chronological order to a standard Gated Recurrent Unit (GRU) \cite{chung2014GRU}. The last hidden state of the GRU is used as the embedding for classification with the MLP.

\vspace{-.2cm}
\subsubsection{RCNN}
We explore two implementations of hybrid recurrent convolutional network to simultaneously model the spatial and temporal structures.
\vspace{.1cm}

\noindent\textit{2.1.3.1 CNN+GRU}\\
Our first implementation successively extracts spatial and temporal embeddings. Each image of the sequence is first independently embedded with a CNN-like network. The resulting sequence of spatial embeddings is then fed to a GRU unit in chronological order. The last hidden state of the GRU is used as a spatio-temporal embedding for classification.

We implement three such models with varying ratios of parameters allocated to the temporal structure. Indeed, we can reduce the number of convolutional kernels and chose a larger hidden state size to increase this ratio  (see Table \ref{tab:params}).

\vspace{.1cm}
\noindent\textit{2.1.3.2 Convolutional Long Short Term Memory(ConvLSTM)}\\
Our second implementation follows the Conv\-LSTM architecture introduced by \cite{xingjian2015convolutional} and implements directly the spatial structure in the recurrent network. It uses image-shaped hidden and cell states, as well as convolutions instead of MLP layers in an LSTM structure \cite{hochreiter1997long}. We refer the reader  to \cite{russwurm2018multi} for more details on this architecture.
\begin{table}[h]
\caption{Summary of the models' parameters.  }
\label{tab:params}
\vspace{-.3cm}
\centering
\scriptsize{
\begin{tabular}{l|cccc}
         & Number of & Hidden State & Temporal  & Total Number  \\
    Model& Kernels   & Size         & Parameter Ratio  & of Parameters \\
        \hline
CNN      & 16 : 32 : 96  & -        &  0   &        $92\,899$         \\
CNN+GRU$_7$  & 16 : 32 : 64  & 64       &  0.7 &    $92\,035$   \\
CNN+GRU$_8$  & 16 : 32 : 36  & 96       &  0.8 &    $93\,807$   \\
CNN+GRU$_9$  & 16 : 16 : 16  & 128      &  0.9 &    $90\,179$   \\
GRU      & -       & 156            &  1   &        $94\,587$   \\
ConvLSTM & 30 : 64     & 64         &  -   &        $95\,353$        
\end{tabular}
}
\end{table}

Table~\ref{tab:params} summarizes the parameters used for the models: the number of kernels for each of the convolutions and the size of the hidden state of the recurrent unit.
We do not consider three dimensional convolutional architectures to cover the spatial and temporal dimension of the data, contrarily to previous studies such as \cite{ji20183d}. Indeed, convolutions are local computations, and hence are not as well-suited for modeling  long term dependencies as recurrent architectures. Additionally, applying  convolutions along the temporal axis assumes that the images of the sequence are regularly sampled in time which is not necessarily the case in practice.

\subsection{Implementation details}

All the models are trained for $50$ epochs with Adam optimizer \cite{kingma2014adam} set with a batch size of $32$ and a learning rate of $10^{-3}$. We use $5$-fold cross-validation: for each fold the dataset is split into train, validation and test set with a 3:1:1 ratio. The epoch achieving the best results on the validation set is used for performance evaluation on the test set.

\begin{figure}[thb]
\begin{minipage}[b]{.48\linewidth}
  \centering
 \centerline{\epsfig{figure=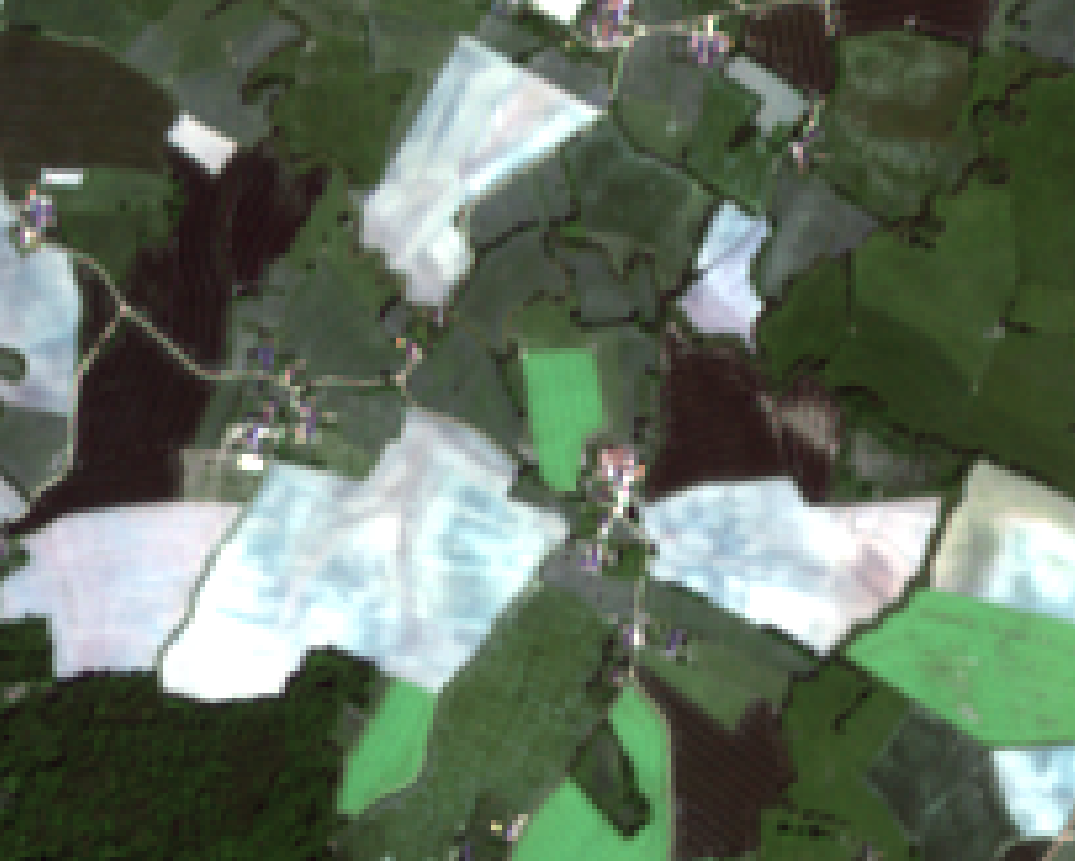,height=2.3cm}}
 \small{
  \centerline{(a) Fragment of tile T31TFM}
  \centerline{of Sentinel-2 TOC product.}}
  \medskip
\end{minipage}
\hfill
\begin{minipage}[b]{0.48\linewidth}
  \centering
 \centerline{\epsfig{figure=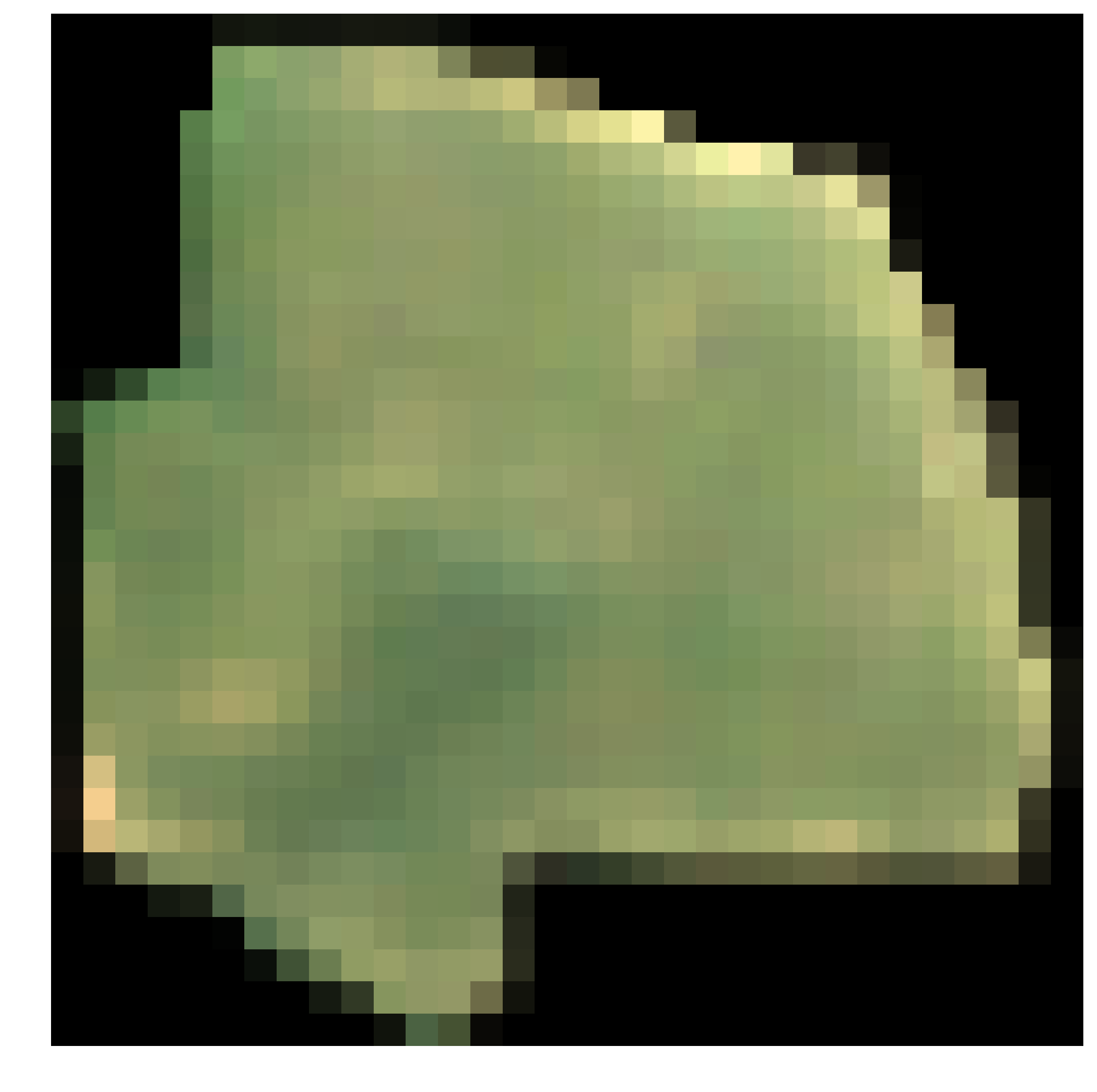,height=2.3cm}}
 \small{
  \centerline{(b) Example of an image patch }
  \centerline{for  one observation of a parcel.}}\medskip
\end{minipage}
\vspace{-0.4cm}
\caption{Example of input image and dataset sample.}
\label{fig:Example}
\vspace{-.5cm}
\end{figure}

\section{Experimental Results}
\label{sec:exp}

\subsection{Dataset}

\label{sec:data}
We test our models using Sentinel-2 multi-spectral image sequences in top-of-canopy (TOC) reflectance (see Figure \ref{fig:Example}(a)). We leave out the atmospheric bands (bands 1, 9, and 10), keeping $10$ spectral bands. 
The area of interest (AOI) corresponds to tile T31TFM in southern France. This tile provides a challenging use case as it presents a high diversity of crops and different terrain conditions. It covers a surface of $12 \, 100$$\:$km$^2$   and contains $199 \, 464$ parcels, each observed at $24$ dates from January to October 2017. 

The values of cloudy pixels are linearly interpolated from the first previous and next available pixel using Orfeo Toolbox \cite{christophe2008orfeo}.
To each parcel we associate a rectangular image patch obtained from the bounding box of its geo-referenced polygon. We set to zero all the pixels outside the parcel as shown on Figure \ref{fig:Example}(b). The patches are resized to $32\times32$ pixels which implies downsampling for only a minority of parcels. 
The dataset is thus comprised of $199 \, 464$ tensors of size $24\times10\times32\times32$. The images are normalized channel-wise for each date individually. 

\begin{figure}[htb]
    \centering
    \includegraphics[width=\linewidth]{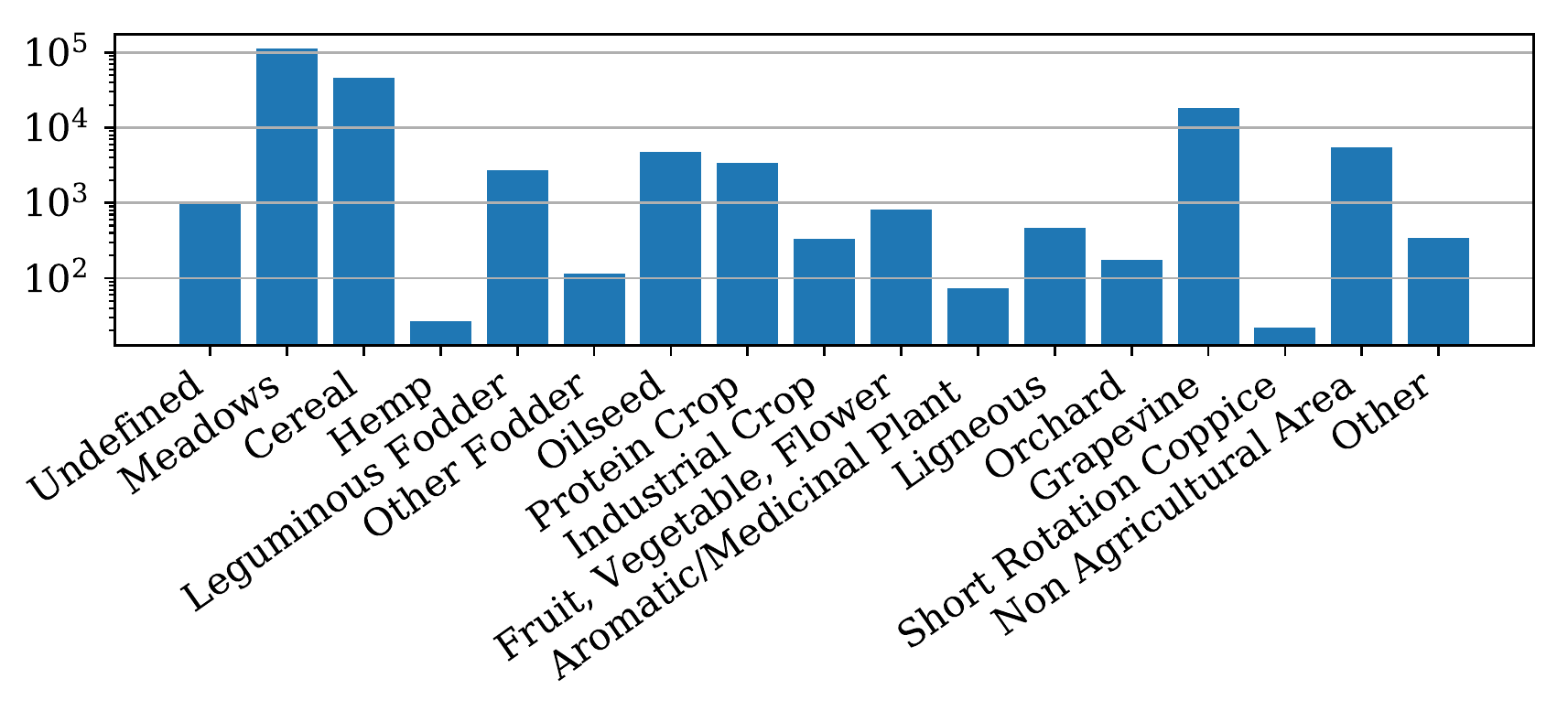}
    \caption{Class breakdown in the AOI. }
    \label{fig:ClassRep}
\end{figure}

We use the French Land Parcel Identification System re\-cords to associate a label to each parcel. The labels are drawn from a comprehensive terminology that covers all observable parcel types and regroups them in 18 high-level classes. Figure~\ref{fig:ClassRep} shows the breakdown of the different classes in the AOI, where meadows, cereals and grapevine are dominant and only the rice class is not represented.

\subsection{Results}
\label{sec:res}
We present the performance of each model in Table~\ref{tab:avg_perf}. Given the high imbalance of the dataset under consideration (see Figure \ref{fig:ClassRep}), we chose the unweighted average class-wise F-score rather than Overall Accuracy (OA) to rank the models' performances. We insist on the fact that all models are designed with approximately the same number of parameters (see Table \ref{tab:params}), so that the differences in performance can only be attributed to the architecture itself,  and not to differences in model complexity.

\begin{table}[h]

\caption{Performance metrics of the different models.}

\centering
\label{tab:avg_perf}
\scriptsize{
\begin{tabular}{l|llll}
Model &  OA          & F-score       & Precision     & Recall\\
\hline
CNN      & 89.5 & 34.9          & 53.0          & 31.4          \\
CNN+GRU$_7$  & 93.4          & 53.2          & 59.7          & 49.9          \\
CNN+GRU$_8$  & 93.7 & 55.1 & 63.3          & 51.6          \\
CNN+GRU$_9$  & \textbf{93.7} & \textbf{55.1}          & \textbf{65.1} & \textbf{51.8} \\
GRU                     & 93.5          & 53.8          & 63.0          & 50.0 \\
ConvLSTM & 93.1 & 49.2 & 64.0 & 44.5 
\end{tabular}
}
\end{table}

Surprisingly, the purely recurrent GRU model outperforms the ConvLSTM and the CNN+GRU$_7$ hybrid models. Only the CNN+GRU$_8$ and CNN+GRU$_9$ models achieve higher overall performance. This shows that extracting both spatial and temporal structures allows for a better classification, only on condition that most of the parameters are allocated to the temporal structure. Only $10\%$ of the parameters seem sufficient for spatial feature extraction with convolutions.

This suggests that the features extracted by RNNs are more discriminative than those extracted by CNNs. Indeed, the purely convolutional model performs significantly worse than its purely recurrent counterpart (by 19 points of F-score).

This performance gap can be explained by the fact that convolutional features are not completely relevant for the problem at hand. CNNs are well-suited for extracting  shape and texture information, and it appears that parcels' shape does not strongly correlate with crop type. Furthermore, the resolution of Sentinel-2 images may not allow to capture rich texture information (see Figure \ref{fig:Example}(b)). This would also explain why the ConvLSTM model --- which relies more on convolutions --- performs slightly worse that the CNN+GRU ones.

Additionally, the results stress the importance of choosing a large enough hidden state to fully leverage the temporal structure of the data. Comparing the GRU and CNN+GRU$_7$ models indicates that  allocating too many parameters to extract convolutional features reduces the performance compared to a model learning on simple hand crafted features with a larger hidden state.

\begin{table}[t]
\caption{Per-class F-score on the test set.}
\label{tab:perclass}
\centering
\footnotesize{
\begin{tabular}{l|cccc}
                         &               &               & CNN+          & Conv          \\
                         & CNN           & GRU           & GRU$_9$        & LSTM          \\
                         \hline
Undefined                & 0.0           & \textbf{45.1}          & 35.4         & 38.2          \\
Meadows                  & 92.9 & 95.8 & \textbf{96.0}          & 96.0         \\
Cereal                   & 93.3 & 97.1          & \textbf{97.5} & 97.5 \\
Hemp                     & 21.6          & 60.5          &       64.4          & \textbf{72.7}          \\
Leguminous Fodder        & 15.6          & 42.9          &\textbf{ 43.8 }         & 34.8          \\
Other Fodder             & 0.0           & 25.5          & \textbf{32.9 }         & 14.8          \\
Oilseed                  & 93.1          & 95.9          & \textbf{96.0}         & 95.1          \\
Protein Crop             & 79.2          & 87.9          & \textbf{89.1 }         & 87.6          \\
Industrial Crop          & 19.6          & 40.3          & \textbf{47.3}          & 23.9          \\
Fruit, Vegetable, Flower & 42.5          & 60.3          & \textbf{63.1}          & 55.1          \\
Aromatic/Medicinal Plant & 0.0           & \textbf{32.8 }         & 28.9          & 8.3           \\
Ligneous                 & 5.5           &\textbf{ 39.8 }         & 38.6         & 32.1          \\
Orchard                  & 0.0           & \textbf{8.0 }          & 6.0         & 2.5           \\
Grapevine                & 81.8          & \textbf{95.1 }         & 94.3         & 92.4          \\
Short Rotation Coppice   & 0.0           & 8.3           & \textbf{18.2 }         & 17.4          \\
Non Agricultural Area    & 46.7          & 54.9          & \textbf{59.1 }        & 55.2          \\
Other                    & 1.2           & 24.0          & \textbf{27.0 }         & 14.1         
\end{tabular}}
\end{table}

We show the per-class performances of the four types of architectures in Table \ref{tab:perclass}. For most classes, the use of convolutional features in the CNN+GRU$_9$ improves the performance compared to the GRU model. Yet some classes such as Grapevine or Ligneous present the opposite behaviour, showing that the relative importance of recurrent and convolutional features is class-dependant.

\begin{table}[h!]
\caption{F-score while trained on the regular image sequences, and with randomly shuffled sequences.}
\label{tab:RnnShuf}
\centering
\scriptsize{
\begin{tabular}{l|ccc}
        & Ordered  & Shuffled  &    $\Delta$    \\
        \hline
ConvLSTM    &       49.2           &         47.9    & -1.3    \\  
CNN+GRU$_7$ &       53.2           &         \textbf{52.0}    & -1.2 \\
CNN+GRU$_9$ &       \textbf{55.1}           &         48.8    &  -6.3  \\
GRU         &       53.8           &         45.3    &  \textbf{-8.5}

\end{tabular}
}
\end{table}

Finally, in order to assess the importance of the temporal structure for the features extracted by the recurrent networks, we re-train several models with randomly shuffled input sequences, such that the temporal structure is lost. Table~\ref{tab:RnnShuf} summarizes the F-scores obtained. 

Time-shuffling impacts all models negatively.  This impact is all the more important as the ratio of temporal parameters is high. Yet, all models still outperform the purely convolutional model. These results suggest that the hidden states of the recurrent units act in two ways: first,  as a memory storing information regardless of the order, and  second, as a model for the temporal evolution of the crops. As our dataset only covers a single year, this chronological evolution is most probably related to the phenology of the crops.

\section{Concluding Remarks}
\label{sec:con}
In this article, we compared the performance of four deep learning architectures extracting spatial, temporal, or spatio-temporal features for crop type mapping with a constraint on the total number of parameters. Our results showed that the best performance is achieved with spatio-temporal features when up to $90\%$ of the parameters are allocated to the temporal structure of the data. 

This suggests that simple convolutional architectures are sufficient to extract expressive features from Sentinel-2 images.
Moreover, this emphasizes the importance of the temporal dimension of Sentinel-2 data to  classify crop types. We showed that RNNs can successfully leverage this structure by acting as a memory combining multiple observations and foremost by taking into account the temporal evolution of the different observations over a year.

In future works, it would be interesting to explore the time-modelling capacity of RNNs in the case of multi-annual observations. Such use cases can present temporal structures of higher order such as the crop rotation patterns studied by \cite{bailly2018crop}. Additionally, the potential of RNNs should be even more valuable with the fully operational revisit time of five days achieved by the Sentinel-2 mission since late 2017. 
Furthermore, we focused here on the temporal and spatial dimensions and treated the spectral dimension with simple concatenation, while learn\-ed spectral features could be beneficial as well. This study should, therefore, be extended to cover the extraction of relevant spectral features with deep learning methods. 

More generally, our results highlight the potential of deep learning models for agricultural parcel classification: all RNN and RCNN models outperform the RF baseline which achie\-ves an average F-score of 36.9 on the same dataset (not shown here). This study  provides a set of guidelines to help in designing bespoke models for the task at hand and contribute to the development of automated crop type mapping at a large scale.

\bibliographystyle{IEEEbib}
\bibliography{igarss_supershort}

\end{document}